\documentclass[prb,twocolumn,eqsecnum]{revtex4}

\begin{document}

\title{Single-mode approximation and effective Chern-Simons
theories \\ 
for quantum Hall systems}
 \author{K. Shizuya}
  \affiliation{Yukawa Institute for Theoretical Physics\\
Kyoto University,~Kyoto 606-8502,~Japan }

\begin{abstract} 
A unified description of elementary and collective excitations in quantum
Hall systems is presented within the single-mode approximation (SMA)
framework, with emphasis on revealing an intimate link with
Chern-Simons theories.  
It is shown that for a wide class of quantum Hall systems the SMA in
general yields, as an effective theory, a variant of the bosonic
Chern-Simons theory.
For single-layer systems the effective theory agrees with the standard
Chern-Simons theory at long wavelengths whereas  substantial deviations
arise for collective excitations in bilayer systems.  
It is suggested, in particular, that Hall-drag experiments would be 
a good place to detect out-of-phase collective excitations inherent 
to bilayer systems.
It is also shown that the intra-Landau-level modes bear a similarity in
structure (though not in scale) to the inter-Landau-level modes, and 
its implications on the composite-fermion and composite-boson theories
are discussed.  
\end{abstract}


\maketitle

\section{Introduction}

The early theoretical study of the fractional quantum Hall
effect~\cite{TSG} (FQHE), based on Laughlin's wave functions,~\cite{L}
revealed that incompressibility is the key character of the quantum Hall
states.
This observation then evolved~\cite{GM,J} into new pictures of the FQHE in
term of electron-flux composites, the  composite bosons or composite
fermions.  There the fractional quantum Hall states are either
visualized as charged superfluids with Bose-condensed composite bosons
in zero magnetic field or mapped to integer-quantum-Hall states of
composite fermions in a reduced magnetic field. 
Chern-Simons (CS) theories, both bosonic~\cite{ZHK,LZ,Z}
and fermionic,~\cite{BW,LF,HLR} realize these composite-particle
descriptions of the FQHE by an expansion around mean field.

There is apparently another stream in the theory of the FQHE,
that enforces the importance of projecting the dynamics onto the
(lowest) Landau level, a key point emphasized in the wave-function
approach.  There are several approximation schemes of this
sort,~\cite{KH,GJ,GMP,HR,MOG,Chak,MZ,Moon} both perturbative and
nonperturbative.
The single-mode approximation (SMA), in particular, is a general
method to study collective excitations in liquid states. 
The SMA equipped with such projection, developed by Girvin,
MacDonald and Platzman,~\cite{GMP} has proved to be a powerful
nonperturbative means to explore quantum Hall systems, 
even better suited than the well-known case of liquid Helium.

The CS and SMA approaches, though equally successful in revealing various
aspects of the FQHE, appear rather independent.
Appealing physical pictures in CS approaches should be contrasted with
the generality in formalism (enforcing Landau-level projection, 
sum rules, etc.) of the SMA theory.  
In certain  cases they even lead to subtle differences.~\cite{MZ}

The purpose of this paper is to present a unified description of
elementary and collective excitations in quantum Hall systems 
within the SMA theory and to uncover an intimate link between 
the SMA and CS theories.
In particular, we show that for a wide class of quantum Hall systems 
the SMA in general yields, as an effective theory, a variant of
the bosonic CS theory.
For single-layer systems the effective theory agrees with the standard
CS theory at long wavelengths whereas  substantial deviations arise for
collective excitations in bilayer systems. 
Such a link between the SMA theory and the composite-boson theory was
earlier noticed indirectly through the response of quantum Hall
states.~\cite{KSdl}  In this paper we establish it directly within the
SMA theory, generalize it to higher-multipole excitations in quantum
Hall systems and discuss its implications  on the composite-fermion
theory as well as the composite-boson theory. 

We present the basic formalism by studying, for single-layer systems,
the cyclotron modes and collective modes in Sec.~II and III,
respectively.
It will be seen that the intra-Landau-level modes bear a similarity in
structure (though not in scale) to the inter-Landau-level modes. 
In Sec.~IV we examine bilayer systems.
Section~VI is devoted to a summary and discussion.

\section{Effective theory -- formalism}

Consider a quantum Hall system described by the Hamiltonian 
\begin{eqnarray}
H_{0}\! &=&\! \int d^{2}{\bf x}\, \Psi^{\dag}(x)\,
 {1\over{2M}} \left({\bf p}+e{\bf A}^{\!B}\right)^{2} \Psi (x)
+ H^{\rm Coul}, \ \ \ \ \ \ \
\label{Hzero}
\end{eqnarray}
where 
${\bf A}^{\!B}=B\,(-x_{2},0)$
supplies a uniform magnetic field  $B_{z}=B>0$ normal to the  
${\bf x}=(x_{1}, x_{2})$ sample plane. 
For definiteness we take a single-layer system with spin-polarized
electrons; the spin and layer degrees of freedom are readily
included.

Out of the electron field $\Psi (x)$ one can form a number of charge
operators.
Actually there are an infinite number of them reflecting the Landau
levels of the electron.
This is made explicit by expanding the electron field  
$\Psi ({\bf x},t)=\sum_{N} \langle {\bf x}|N\rangle\, \Psi_{n}(y_{0},t)$ 
in terms of the Landau levels 
$|N\rangle = |n,y_{0}\rangle$ of a freely orbiting electron
of energy $\omega_{c} (n+{\scriptstyle {1\over2}})$ with 
$n = 0,1,2,\cdots$, and $y_{0}=\ell^{2}\,p_{x_{1}}$, where
$\omega_{c} \equiv eB/M$ and ${\ell}\equiv 1/\sqrt{eB}$; 
we frequently set $\ell \rightarrow 1$ below. 
The charge operator $\rho_{\bf p}= \int d^{2}{\bf x}\, e^{-i{\bf
p\cdot x}}\,\Psi^{\dag}\Psi$ then reads~\cite{GJ,GMP,KS}
\begin{eqnarray} 
\rho_{\bf p}&=& \sum_{m,n=0}^{\infty} F^{mn}({\bf p})\, R^{(mn)}_{\bf p},
\nonumber \\
R^{(mn)}_{\bf p}\! &=& \!
\int dy_{0}\ \Psi_{m}^{\dag}(y_{0},t)\,
e^{-{1\over{4}} \ell^{2} {\bf p}^{2}} e^{-i{\bf p\cdot r}}\,
\Psi_{n}(y_{0},t).\ \ \ \ 
\end{eqnarray}
Here ${\bf r} \equiv (r_{1},r_{2}) = (i\ell^{2} \partial/\partial y_{0},
y_{0})$ stands for the center coordinates
with uncertainty $[r_{1}, r_{2}]=i\ell^{2}$, and  
\begin{eqnarray}
F^{mn}({\bf p}) \!
&=& \langle m| e^{-i(p/\!\sqrt{2})Z^{\dag}} 
e^{-i(p^{\dag}/\!\sqrt{2}) Z} |n\rangle  \ \ \ \
\label{Fp}
\end{eqnarray}
with $[Z, Z^{\dag}]=1$ and $Z^{\dag}Z|n\rangle  =n |n\rangle$;
$p=p_{2}+ ip_{1}$.
In particular, $F^{mn}({\bf p})\!=\! 
\sqrt{n!/m!}\,(-ip/\sqrt{2})^{m-n}\, L^{(m-n)}_{n} ({{\bf p}^{2}/2})$ 
for $m \ge n$;
$F^{00}({\bf p}) =1$, $F^{n0}({\bf p}) = (-ip/\sqrt{2})^{n}/\sqrt{n!}$,
and $F^{0n}({\bf p}) = (-ip^{\dag}/\sqrt{2})^{n}/\sqrt{n!}$.
The fields $\Psi_{n}(y_{0},t)$ obey the canonical anticommutation
relations $\{\Psi_{m}(y_{0},t),\Psi_{n}^{\dag}(y'_{0},t)\}  
= \delta_{mn}\, \delta (y_{0}-y'_{0})$.  
The underlying Landau-level structure is now encoded in the 
$U_{\infty}$ or $W_{\infty}$ algebra~\cite{GMP} obeyed by the charges
$R^{(mn)}_{\bf p}$:
\begin{eqnarray}
[R^{(m m')}_{\bf k}, R^{(n n')}_{\bf p}] 
\!&=&\! \delta^{m'n}e^{{1\over{2}} k^{\dag} p}
R^{(m n')}_{\bf k+p} - \delta^{n' m}e^{{1\over{2}} p^{\dag} k}
R^{(n m')}_{\bf k+p}. \nonumber\\
\label{chargealgebra}
\end{eqnarray}

The  charges $\rho^{(mn)}_{\bf p}= F^{mn}({\bf p})\, R^{(mn)}_{\bf p}$ 
generate inter- and intra-Landau-level excitations for $m\not=n$ and
$m=n$, respectively. 
Our task in this section is to study such excitations, both elementary
and collective, over a quantum Hall state by means of the
single-mode approximation~\cite{GMP,MOG} (SMA).
Let $|G\rangle$ denote an exact quantum Hall (i.e., incompressible)
ground state of the Hamiltonian $H_{0}$ with uniform density $\rho_{0}$.
In the SMA one represents the excitation modes as 
$|\phi^{\alpha}_{\bf k}\rangle \sim \rho^{\alpha}_{\bf k}|G\rangle$ with
$\alpha = (m,n)$, and regards their normalization 
\begin{eqnarray} 
s^{\alpha}({\bf k}) &=& (1/N_{e})\, 
\langle G|(\rho^{\alpha}_{\bf k})^{\dag} \rho^{\alpha}_{\bf k}|G\rangle,
\label{skalpha}
\end{eqnarray}
called the static structure factors, as the basic quantity;
$N_{e}$ stands for the total electron number.
Saturating the $f$-sum rule or the oscillator strength
\begin{equation}
f^{\alpha}({\bf k}) = (1/N_{e})\, \langle G|\,
(\rho^{\alpha}_{\bf k})^{\dag}\, [H_{0} , \rho^{\alpha}_{\bf k}]\, 
|G\rangle,
\end{equation}
calculable by use of the charge algebra~(\ref{chargealgebra}),
with the single mode $|\phi^{\alpha}_{\bf k}\rangle$ then yields 
the SMA excitation spectrum $\epsilon^{\alpha}_{\bf k} 
=f^{\alpha}({\bf k})/s^{\alpha}({\bf k})$.  
The spectrum is determined once $s^{\alpha}({\bf k})$ is known.

The SMA sets up one-to-one correspondence between the excitations and
Landau-level charges $\rho^{(mn)}_{\bf p}$.  
The assumption of single-mode dominance is far from obvious but
exact-diagonalization studies of small systems generally suggest that
it is a good approximation for the lowest-lying collective
modes.~\cite{GMP}  We therefore pursue the SMA here.

Let us now try to construct an effective theory realizing the SMA
description of excitations.
The correspondence noted above suggests us to use the
technique~\cite{Moon} of nonlinear realization of the $W_{\infty}$
algebra for this purpose.  
First let $\Psi^{\rm cl}_{0}(y_{0},t)$ denote a classical configuration
or the ground-state configuration, with pertinent correlations
characterized by (a set of) static structure factors 
$s^{\alpha}({\bf k})$.  We then write the electron field $\Psi$ in
the form of a small variation in phase from $\Psi^{\rm cl}$,
\begin{eqnarray}
\Psi\sim  \exp [-i\sum_{\alpha}\sum_{\bf p}
\theta_{\bf -p}^{\alpha} T^{\alpha}_{\bf p}]\,\Psi^{\rm cl}.
\label{transformpsi}
\end{eqnarray} 
Here $\theta_{\bf -p}^{\alpha}=\theta_{\bf -p}^{(mn)}$ 
with $(\theta_{\bf p}^{(mn)})^{\dag}= \theta_{\bf -p}^{(nm)}$ stand for
local phase variations;
$T^{\alpha}_{\bf p} = F^{mn}({\bf p})\,
e^{-{1\over{4}}\, {\bf p}^{2}} e^{-i{\bf p\cdot r}}$.
Rewriting the Lagrangian in favor of $\Psi^{\rm cl}$ and $\theta$, and
replacing the products of $(\Psi^{\rm cl})^{\dag}$ and $\Psi^{\rm cl}$ by
the structure factors then yields an effective Lagrangian for the
excitation modes $\theta_{\bf p}^{\alpha}$.
  
For such transcription it is convenient to express
Eq.~(\ref{transformpsi}) in operator form 
\begin{eqnarray}
&&\Psi = {\cal P}^{-1}\,\Psi^{\rm cl}\,{\cal P},
\nonumber\\
&&{\cal P} = e^{-i\theta \cdot \rho^{\rm cl}}, \
\theta \!\cdot\! \rho^{\rm cl} \equiv  \sum_{\bf p}\sum_{\alpha}
\theta_{\bf -p}^{\alpha}  (\rho^{\alpha}_{\bf p})^{\rm cl}, {\rm etc.}
\label{psivspsicl}
\end{eqnarray} 
Here $(\rho^{\alpha}_{\bf p})^{\rm cl}$ stand for $\rho^{\alpha}_{\bf p}$
with $\Psi$ replaced by $\Psi^{\rm cl}$, and obey the same charge 
algebra as $\rho^{\alpha}_{\bf p}$. Repeated use of the algebra then
enables one to express  
$\rho^{\alpha}_{\bf p}= {\cal P}\,(\rho^{\alpha}_{\bf p})^{\rm cl}\, 
{\cal P}^{-1}$ 
in powers of $(\rho^{\alpha}_{\bf p})^{\rm cl}$.

Substituting Eq.~(\ref{transformpsi}) into the Lagrangian
$L= \int d^{2}{\bf x}\, \Psi^{\dag} i\partial_{t} \Psi - H_{0}$
yields a Berry's phase~\cite{Berry} contribution
$(\Psi^{\rm cl})^{\dag}\,e^{i\theta\cdot T} (i \partial_{t}
e^{-i\theta\cdot T})\Psi^{\rm cl}$, which, by use of the charge
algebra~(\ref{chargealgebra}), is cast in the compact form
\begin{eqnarray}
e^{i\theta\cdot\rho^{\rm cl}}\, i\partial_{t} 
e^{-i\theta\cdot\rho^{\rm cl}} = \dot{\theta} \!\cdot\! \rho^{\rm cl} +
{i\over{2}}\, [\theta \!\cdot\! \rho^{\rm cl},
\dot{\theta}\!\cdot\! \rho^{\rm cl}] +\cdots,
\end{eqnarray}
where $\partial_{t}$ acts on $\theta$;
$\dot{\theta}=\partial_{t}\theta$.
The Lagrangian then reads 
\begin{eqnarray}
L= e^{i\theta\cdot\rho^{\rm cl}}\, 
(i\partial_{t} - H_{0}^{\rm cl} )\,  e^{-i\theta\cdot\rho^{\rm cl}}
\label{Ltheta}
\end{eqnarray}
where $H_{0}^{\rm cl}$ stands for $H_{0}$ with 
$\Psi \rightarrow \Psi^{\rm cl}$.
We shall from now on exclusively handle $\Psi^{\rm cl}$ and
$(\rho^{\alpha}_{\bf p})^{\rm cl}$, and suppress the suffix 
$\lq\lq{\rm cl}"$ unless confusion arises.

Suppose now that $|G\rangle$ consists predominantly of the
lowest-Landau-level $(n=0)$ components, with the filling factor 
$\nu = 2\pi \ell^{2} \rho_{0}<1$. Taking the expectation value of
Eq.~(\ref{Ltheta}) then yields the effective Lagrangian
\begin{eqnarray}
\langle L\rangle 
= \rho_{0} \sum_{\bf k} \sum_{n=1}^{\infty} 
\theta^{(0n)}_{\bf k} \left[ i s^{(n0)}({\bf k})\partial_{t} + 
f^{(n0)}({\bf k}) \right] \theta^{(n0)}_{\bf -k}\ \ \ \ \ 
\label{Lexp}
\end{eqnarray}
to $O(\theta^{2})$, apart from total divergences; $\langle \cdots
\rangle \equiv \langle G|\cdots |G\rangle$ for short.
Here $s^{(n0)}({\bf k}) = (1/N_{e}) 
\langle G|\, \rho^{(0n)}_{\bf -k}\, \rho^{(n0)}_{\bf k}|G \rangle$ 
are the structure factors  and $f^{(n0)}({\bf k})$ are 
the associated oscillator strengths; 
$\rho_{0}=N_{e}/\Omega$ with the total area $\Omega$. 
Actually, noting that 
$\langle R^{(00)}_{\bf k} \rangle = \rho_{0}\,\delta_{\bf k,0}$, 
and $\langle R^{(0n)}_{\bf k} R^{(n0)}_{\bf p}\rangle =
\rho_{0}\,e^{-{1\over{2}}\,{\bf k}^{2}}\, \delta_{\bf k+p,0}$
with $\delta_{\bf k,0}=(2\pi)^{2}\, \delta^{2} ({\bf k})$,
one finds that  
\begin{eqnarray}
s^{(n0)}({\bf k}) &=& (1/n!)\, (\ell^{2}\,{\bf k}^{2}/2)^{n}\,
e^{-{1\over{2}}\, \ell^{2}\, {\bf k}^{2}}, \nonumber \\
f^{(n0)}({\bf k}) &=& n\omega_{c}\, s^{(n0)}({\bf k}) 
+ O(H^{\rm Coul}).
\end{eqnarray}
for $ n\ge 1$.
The Lagrangian~(\ref{Ltheta}) tells us that $\theta^{(n0)}$ are
canonically conjugate to 
$\theta^{(0n)}=(\theta^{(n0)})^{\dag}$ for $n\ge 1$. 
They describe the $(0\rightarrow n)$ cyclotron modes with the spectrum
$\epsilon^{(n0)}_{\bf k}= f^{(n0)}({\bf k})/s^{(n0)}({\bf k}) \approx
n\omega_{c}$; physically  they represent neutral exciton
excitations~\cite{KH} formed of a hole in the zero-th level and 
an electron in the $n$th level. 
Note here that the intra-Landau-level mode $\theta^{(00)}$, which
finds no canonical conjugate, requires a separate analysis, 
which will be given in Sec.~III.

Let us now generalize the framework by including coupling 
to weak external electromagnetic potentials
$A_{\mu}(x)=(A_{1},A_{2}, A_{0})$.
Replacing  the kinetic term in Eq.~(\ref{Hzero}) with 
$(1/2M)({\bf p} + e{\bf A}^{\!B} + e{\bf A})^{2} + eA_{0}$ yields 
the Hamiltonian $H=H_{0} +U$ with
\begin{eqnarray}
U &=& \sum_{\bf p}\sum_{m,n}U^{mn}_{\bf -p} R^{(mn)}_{\bf p}
\nonumber\\
U^{mn}_{\bf -p}&=& i\ell\omega_{c} \Big\{ 
A_{\bf -p} \sqrt{m}F^{m-1,n}({\bf p}) 
- A^{\dag}_{\bf -p} \sqrt{n}F^{m,n-1}({\bf p}) \Big\}
\nonumber\\
&&+ \tilde{\chi}_{\bf -p} F^{mn}({\bf p}) ,
\end{eqnarray}
where  $A(x)= (A_{2} +iA_{1})/\sqrt{2}$ and 
$A^{\dag} = (A_{2} -iA_{1})/\sqrt{2}$;
$\tilde{\chi}= A_{0} + (1/2M)\, (A_{12} + A_{k}^{2})$,
$A_{12} =\partial_{1}A_{2} -\partial_{2}A_{1}$;
$\tilde{\chi}_{\bf p}, A_{\bf p}$, etc., stand for the Fourier
transforms (with obvious time dependence suppressed); 
the electric charge $e$ has been suppressed by rescaling 
$eA_{\mu} \rightarrow A_{\mu}$.

The $U$ turns into the effective interaction
${\cal P} U{\cal P}^{-1} = U + i[\theta\!\cdot\!\rho, U] +\cdots$.
It suffices to calculate only the $O(A\theta)$ coupling to derive
the electromagnetic response to $O(A^{2})$ eventually.
To simplify the result it is convenient to express $\theta^{(n0)}$ 
in terms of real fields $\eta^{(n)}(x)$ and $\xi^{(n)}(x)$,
\begin{eqnarray}
&& \gamma_{\bf k}\, \theta^{(n0)}_{\bf k}  = \xi^{(n)}_{\bf k}
+i\eta^{(n)}_{\bf k}, \nonumber\\
&& \gamma_{\bf k}= {1\over{\sqrt{n!}}}\,
\Big( {1\over{2}}\,\ell^{2}{\bf k}^{2}\Big)^{(n-1)/2}\, 
e^{-{1\over{4}}\ell^{2} {\bf k}^{2}},
\end{eqnarray}
and to pass from the Fourier space to real ${\bf x}$ space.
The effective Lagrangian to $O(A\theta)$ is cast in the form  
$\langle L\rangle = \int d^{2}{\bf x} ({\cal L}_{A} 
+ {\cal L}_{A\theta})$  with
\begin{eqnarray}
{\cal L}_{A} &=& -\rho_{0}
\Big( A_{0} + {1\over{2M}}\, A_{j}^{2}\, \Big), \\
{\cal L}_{\theta A} &=& \rho_{0} \Big[ 
 2\,\eta\,s\, (\dot{\xi} - \gamma\, \chi)  
-(\eta f \eta+ \xi f \xi ) \nonumber\\
&&\ \ \ \ \ \ \  - \kappa \ell^{2} \xi\,\gamma \partial_{j}A_{j}
+\kappa \ell^{2} \eta\,\gamma A_{12}   
\Big]. 
\label{LthetaA}
\end{eqnarray}
Here $\chi= A_{0} + (1/2M)\, A_{12}$, $\kappa =  n\omega_{c}$ and 
\begin{eqnarray}
&&s\equiv s^{(n0)}({\bf k})/\gamma^{2} = \ell^{2}{\bf k}^{2}/2,
\nonumber\\ 
&&f\equiv f^{(n0)}({\bf k}) /\gamma^{2} = n\omega_{c}\, 
\ell^{2}{\bf k}^{2}/2, 
\label{reducedsf}
\end{eqnarray}
with  ${\bf k} = -i\nabla$ understood and $\gamma =\gamma_{\bf k}$.
The ${\cal L}_{A}$ comes from $-\langle U\rangle$ and 
${\cal L}_{\theta A}$ involves the $O(\theta^{2})$ term of
Eq.~(\ref{Lexp}) as well. For conciseness explicit reference to the mode
index $n$ has been suppressed in the above; remember that a summation
over all modes is implied in ${\cal L}_{\theta A}$.

It is a straightforward but strenuous task to calculate the 
$O(H^{\rm Coul})$ portion~\cite{MOG} of $f^{(n0)}({\bf k})$, 
left out from $f$ in Eq.~(\ref{reducedsf}).  
An alternative and simpler way, that works to extract the leading
long-wavelength part of it, is to first linearize the Coulomb
interaction
${1\over{2}}\sum_{\bf k} V_{\bf k}\rho_{\bf -k} \rho_{\bf k}$ by
use of a Hubbard-Stratonovich field $\phi$. 
The net effect is to replace $\chi$ by $\chi + \phi$ in
Eq.~(\ref{LthetaA}). Eliminating $\phi$ then yields the effective
interaction
\begin{eqnarray}
 {\cal L}^{\rm Coul} = - 2\rho_{0}^{2}\,
 \eta\,s\gamma\, V_{\bf k}\,s\gamma\,\eta.
\end{eqnarray}

The $(0\rightarrow n)$ exciton modes are now described by the
effective Lagrangian
${\cal L}_{A} + {\cal L}_{A\theta} + {\cal L}^{\rm Coul}$.  
Let us rewrite it in a more suggestive form.  We consider the
$(0\rightarrow1)$ mode first and set $\gamma \rightarrow 1$.
For generality and later use we leave the normalization of the
reduced factors,
$\hat{s}\equiv s/(\ell^{2}{\bf k}^{2}/2)$ and 
$\hat{f}\equiv f/(\ell^{2}{\bf k}^{2}/2)$, arbitrary and denote 
the spectrum $\epsilon =f/s =\hat{f}/\hat{s} $ [although
$\hat{s}=1$ and $\hat{f}=\epsilon =n\omega_{c}$ in the present case].
Let us combine the $\xi^{2}, \xi \partial_{j}A_{j}$ and $\eta A_{12}$
terms to form a complete square so that
\begin{eqnarray}
&&{\cal L}_{A\theta}+{\cal L}^{\rm Coul} \nonumber\\
&&=r(\dot{\xi} - A_{0}) 
 -{1\over{2}}\, \rho_{0}\ell^{2} \hat{f}\,\Big(\partial_{j}\xi-
{\kappa\over{\hat{f}}}\, A_{j}\! + {\delta\over{2\rho_{0} s}}\, 
\epsilon_{jk} \partial_{k} r \Big)^{2} \nonumber\\
&&-{1\over{2}}\, r\, (V_{\bf k} + \triangle)\, r
+ {1\over{2}}\,\rho_{0}\ell^{2}\kappa^{2}
A_{j}\, {1\over{\hat{f}}}A_{j},
\label{rwtLAtheta} 
\end{eqnarray}
where we have set $r=2\rho_{0} s\,\eta$ and 
$\triangle = (1 - \delta^{2} )\epsilon /(2\rho_{0}s)$ with
$\delta = 1- s\omega_{c}/\kappa$; $\epsilon_{12} = -\epsilon_{21}=1$.

Note that for gauge invariance it is necessary to have
$\kappa/\hat{f}=1$  or $f={1\over{2}}\, \kappa \ell^{2}{\bf k}^{2}
+O({\bf k}^{4})$,  which is satisfied in the present case. 
The last term in Eq.~(\ref{rwtLAtheta}), though spoiling gauge
invariance, is harmless. For the $(0 \rightarrow n)$ mode it reads
$\propto {1\over{2}}\,\rho_{0}\ell^{2} n
\omega_{c} A_{j}\, (\gamma_{\bf k})^{2}A_{j}$, which, when summed
over $n=1,2,\cdots$, amounts to 
${1\over{2}}\,\rho_{0}\ell^{2} \omega_{c} A_{j}^{2}$. 
Thus this term combines with the $A^{2}$ term in ${\cal L}_{A}$ to
vanish exactly.

The $(1/s)\, \epsilon_{jk} \partial_{k}\, r$ term in
Eq.~(\ref{rwtLAtheta}) can be disentangled by use of a vector field
$c_{\mu}$, thereby yielding an effective Chern-Simons theory
described by the Lagrangian
\begin{eqnarray}
{\cal L}_{c} &=&  r(\dot{\xi} - A_{0}- c_{0}) 
-{1\over{2}}\, \rho_{0}\ell^{2}\, \hat{s}\,
\epsilon\,(\partial_{j}\xi- A_{j} - c_{j})^{2}\nonumber\\
&& -{1\over{2}}\, r\, (V_{\bf k} +\triangle )\, r  
-{1\over{2}}\,  
c_{\mu}\alpha \epsilon^{\mu\nu\rho}\partial_{\nu}c_{\rho} 
\label{Leffcs}
\end{eqnarray}
with $\alpha = \rho_{0} \ell^{2}\kappa/(\epsilon \delta)$; 
$\epsilon^{012} =1$. [One may practically set 
$\delta = 1 + O(\ell^{2}{\bf k}^{2}) \rightarrow 1$ and 
$\alpha \rightarrow \rho_{0} \ell^{2} = \nu /(2\pi)$ here. 
For the full Lagrangian, include in ${\cal L}_{c}$ 
the $- \rho_{0}A_{0}$ term coming from ${\cal L}_{A}$.]  
The equivalence of ${\cal L}_{c}$ to 
${\cal L}_{A\theta}+{\cal L}^{\rm Coul}$ is immediately seen 
in the Coulomb gauge $\partial_{k}c_{k}=0$.

For the Laughlin sequence $\nu =1/3,1/5, \cdots$ 
this effective theory precisely agrees with the standard CS
theory~\cite{ZHK} with the composite-boson field 
$\phi_{\rm cb} (x) = \sqrt{\rho_{0} + r}\, e^{-i\xi}$ 
expanded to second order in $r$ and $\xi$ around the mean field, 
except for the 
$\triangle \approx \epsilon\, \omega_{c}/(\rho_{0} \kappa)$ term which
in the latter reads $\ell^{2}{\bf k}^{2}\omega_{c}/(4\rho_{0})$. 
This $\triangle$ term is less important than the Coulomb term 
$V_{\bf k} \sim 1/|{\bf k}|$ at long wavelengths.

The filling factor $\nu < 1$ is not fixed within the present approach. 
The effective CS theory~(\ref{Leffcs}) makes sense for general $\nu$ 
as long as (i) the state $|G\rangle$ is incompressible and 
(ii) the SMA is applicable (approximately).

It is possible to cast the theory into an equivalent 
dual-field~\cite{LZ} form, suited for discussing the dynamics of
vortices.
We first use a vector field $j_{k}$ to linearize 
the complete square term in Eq.~(\ref{rwtLAtheta}) so that
$j_{k}\,(\partial_{k}\xi + \cdots) + (1/2\rho_{0})\,
j_{k}(1/\hat{f})j_{k}$ and then eliminate $\xi$.
The resulting conservation law $\partial_{\mu}j_{\mu}=0$ with 
$j_{0}\equiv r$ allows one to set 
$j_{\mu}=\epsilon^{\mu \nu \lambda}\partial_{\nu}b_{\lambda}$ 
with a three-vector field $b_{\mu}=(b_{k},b_{0})$.  
Substituting this back into Eq.~(\ref{rwtLAtheta}) then yields 
an equivalent theory described by the Lagrangian
\begin{eqnarray}
{\cal L}_{b} &=& -A_{\mu}  \epsilon^{\mu
\nu\rho}\partial_{\nu} b_{\rho} +  {1\over{2\rho_{0}
\ell^{2}}}\, b_{\mu} {\delta\over{\hat{s}}} \,  
\epsilon^{\mu \nu\rho}\partial_{\nu} b_{\rho} \nonumber\\
&&+ {1\over{2\rho_{0}\ell^{2}}}\,b_{k0}
{1\over{\hat{s}\, \epsilon}}\, b_{k0}
-{1\over{2}}\, b_{12}\, (V_{\bf k} +\triangle)\,b_{12}.\ \ \
\label{Lb}
\end{eqnarray}
The third term is obtained with the aid of the formula
\begin{eqnarray}
j_{k}\epsilon_{ki}\partial_{i}j_{0} 
 &=& -{1\over{2}}\,b_{\mu} \nabla^{2}\,\epsilon^{\mu \nu
\rho}\partial_{\nu}b_{\rho} + {\rm total\  div.}  
\end{eqnarray}

From this ${\cal L}_{b}$ one can calculate an electromagnetic
response of the form $S[A] =\int d^{3}x\, {\cal L}[A]$ with
\begin{eqnarray}
{\cal L}[A] &=& {1\over{2}}\, \Big\{
- A_{\mu}\,\delta\,{\cal D}\epsilon^{\mu\nu\rho}\partial_{\nu}A_{\rho} 
+ A_{k0}{1\over{\epsilon}}\, {\cal D} A_{k0} \nonumber\\ 
&& \ \ \ \ \ \ 
-A_{12} \Sigma\, {\cal D}A_{12} \Big\}, 
\label{emresponse}\\
{\cal D}&=&  \rho_{0}\ell^{2}\,\hat{s}\,
{\epsilon^{2}\over{E_{\bf k}^{2} -\omega^{2} }},
\end{eqnarray}
where $\omega =i\partial_{t}$ and 
$\Sigma = \rho_{0}\ell^{2}\,\hat{s}\, (V_{\bf k} +\triangle)$, 
with the excitation spectrum given by
\begin{eqnarray}
E_{\bf k}^{2}&\approx& \epsilon^{2} + 2\rho_{0}\epsilon s\, V_{\bf k}. 
\end{eqnarray}
Note that the Coulomb interaction hardly affects the long-wavelength
response.

So far we have focused on the $(0\rightarrow 1)$ cyclotron mode. 
For the higher modes one may simply set 
$A_{\mu} \rightarrow \gamma_{\bf k}\,A_{\mu}$,
$V_{\bf k} \rightarrow \gamma_{\bf k}\, V_{\bf k}\,\gamma_{\bf k}$ 
and $\epsilon =\kappa \rightarrow n\omega_{c}$ in 
${\cal L}_{c},$ ${\cal L}_{b}$ and ${\cal L}[A]$ in the above. 
It is seen that the $(0\rightarrow n)$ exciton modes lead to the
density-density response of 
$\langle \rho \rho \rangle \sim ({\bf k}^{2})^{n}$ for $n \ge 1$.
Thus only the $(0\rightarrow 1)$ mode is dipole-active, 
i.e., sensitive to long-wavelength probes.

A link between the SMA theory and CS theories was noticed
earlier~\cite{KSdl} for dipole-active excitations, indirectly through the
response of quantum Hall systems.  The advantage of the present
approach, we remark, lies in establishing such a link directly within the
SMA theory and, moreover, allowing one to study dipole-inactive
(higher-multipole) excitations equally well.

\section{Intra-Landau-level excitations}

In this section we introduce a variational principle for handling
intra-Landau-level excitations. Experimentally collective excitations
over the $\nu =1/3$ Laughlin state have been observed  
by inelastic light scattering.~\cite{PDPW}  
Correspondingly we here focus on excitations within the lowest 
Landau level, and denote the relevant $\theta_{\bf k}^{(00)}$ mode 
by $\bar{\theta}_{\bf k}$ and the charge operator 
$\rho^{(00)}_{\bf p}$ by $\bar{\rho}_{\bf p}$ for short;
$(\bar{\theta}_{\bf k})^{\dag} =\bar{\theta}_{\bf -k}$ so that 
$\bar{\theta}(x)$ is a real field.

Let us denote the Lagrangian~(\ref{Ltheta}) anew as
$L_{\theta} = e^{i\theta\cdot\rho}\,  
(i\partial_{t} - H_{0})\,  e^{-i\theta\cdot\rho}$ with
$\theta \cdot\!\rho \rightarrow \sum_{\bf p} \bar{\theta}_{\bf -p}  
\bar{\rho}_{\bf p}$ (and the superscript "cl" suppressed as before).
This $L_{\theta}$ involves no canonical momentum conjugate 
to $\bar{\theta}(x)$. We try to supply it through a local
variation in amplitude of the field. There are a number of ways to
achieve this, but not all of them embody the SMA formalism.
The variational ansatz we adopt is to consider 
\begin{eqnarray}
L_{\lambda\theta} = \langle G|
e^{\lambda\cdot\rho} L_{\theta}\, e^{\lambda\cdot\rho}|G \rangle 
=\langle G| \{\lambda\! \cdot \! \rho, \dot{\theta} \!\cdot\!
\rho\}  +\cdots |G \rangle,
\label{Lthetalambda}
\end{eqnarray}
with $\lambda \!\cdot\! \rho = \sum_{\bf k}
\lambda_{\bf -k}  \bar{\rho}_{\bf k}$, where $\lambda_{\bf k}$ 
[or the real field $\lambda (x)$]
denotes a local amplitude modulation.

Some care is needed here. The intra-Landau-level modes are governed 
by the Coulomb interaction and depend critically on the definition 
of the Landau levels.  
In order to derive a gauge-invariant result, it turns out
necessary to define the Landau levels properly with the external
potentials $A_{\mu}$ taken into account.
We thus first make a unitary transformation
$\Psi_{m}(y_{0},t) \rightarrow \Psi'_{n}(y_{0},t)$ so that 
the one-body Hamiltonian becomes diagonal in Landau-level indices.  
Such a procedure of projection was developed earlier.~\cite{KSdl}  
The one-body Hamiltonian thereby  acquires terms of $O(A^{2})$, 
which precisely agree with the electromagnetic
response~(\ref{emresponse}) due the cyclotron modes;
thus, no explicit account of them is needed here. 
The relevant one-body Hamiltonian to $O(A)$ reads
\begin{eqnarray}
\bar{H}_{1}=\sum_{\bf p}\Big({1\over{2}}\,\omega_{c} \delta_{\bf p,0}
+ \chi_{\bf p} \Big) \,
\bar{\rho}_{\bf -p},
\end{eqnarray}
with $\chi = A_{0} + (1/2M)\, A_{12}$.
Here $\bar{\rho}_{\bf p}$ are the charge operators defined in terms of
the transformed field $\Psi'_{n=0}(y_{0},t)$; we suppress this
specification because they obey the same algebra as
$R^{(00)}_{\bf p}$ in Eq.~(\ref{chargealgebra}).

An important consequence of projection is that the projected Coulomb
interaction~\cite{fnbg}  
\begin{eqnarray}
\bar{H}^{\rm Coul} &=& {1\over{2}}\, \sum_{\bf p} V_{\bf p}\,
\bar{\rho}_{\bf -p}\,\bar{\rho}_{\bf p}
+ \triangle H^{\rm Coul},
\end{eqnarray}
acquires a field-dependent piece~\cite{KSdl} (which actually comes
from the gauge invariance of the projected charges)
\begin{eqnarray}
\triangle H^{\rm Coul} &=&  {1\over{2}}\sum_{\bf p,k} V_{\bf p}\, 
 u_{\bf p,k}\, \{ \bar{\rho}_{\bf -p},
\bar{\rho}_{\bf p - k} \},
\label{triHC}
\end{eqnarray}
where
\begin{eqnarray}
u_{\bf p,k} 
= i\!\!  &\Big[\!\!\!& {\bf p\! \times\! A} 
- {1\over{4}}\, \{ {\bf p\! \cdot\! k}\, ({\bf p\! \times\! A}) 
+  ({\bf p\! \times\! k})\,{\bf p\! \cdot\! A} \} 
\nonumber\\
&& - {1\over{2}}\,{\bf p\! \cdot\! k}\, ({\bf k\! \times\! A}) 
 + O({\bf p}^{3}{\bf k}^{2} {\bf A}) \Big], 
\end{eqnarray}
apart from terms of $O(1/\omega_{c})$; 
${\bf p \times A} = \epsilon_{ij}p_{i}(A_{j})_{\bf k}$,  
${\bf p \cdot A} = p_{i}(A_{i})_{\bf k}$, etc. 
In Eq.~(\ref{triHC}) we have retained only $O(A)$ corrections, 
relevant to our discussion below.

Let us first turn $A_{\mu}$ off.
Then the ground state $|G\rangle$ is an eigenstate of the Coulomb
interaction,
$H_{0}|G\rangle =E_{0}|G\rangle$ with $H_{0}=\bar{H}^{\rm Coul} +$
quenched kinetic terms.  We set $E_{0}=0$ by shifting the zero of energy.
This is an important step since terms like 
$\langle \{ \lambda \cdot \rho, H_{0}\} \rangle \sim \lambda_{\bf k=0}$,
which drive a constant shift in $\lambda$, thereby vanish.
[The necessity of this shift in energy is consistent with the fact that
the oscillator strength $f({\bf k})$ involves only excitation energies
measured relative to $E_{0}$.]
The effective Lagrangian then takes the form 
$L_{\lambda\theta}= \int d^{2}{\bf x}\, {\cal L}_{0}$ with
\begin{eqnarray}
{\cal L}_{0}
=\rho_{0}\, \Big[2\, \lambda\, \bar{s} \, \dot{\bar{\theta}} - 
 (\lambda\,\bar{f}\, \lambda+ \bar{\theta}\,\bar{f}\, \bar{\theta} )
\Big].
\label{LDL}
\end{eqnarray}
This shows that $\lambda$ serves as a canonical conjugate of
$\bar{\theta}$ and that $\bar{\theta}$ describes excitations with 
the spectrum $\epsilon = \bar{f}/\bar{s}$.

In general, the intra-Landau-level mode is dipole-inactive~\cite{GMP} 
and the structure factor 
$\bar{s}({\bf k}) = (1/N_{e})\, \langle \bar{\rho}_{\bf -k}\,
\bar{\rho}_{\bf k}\, \rangle$ starts with $O({\bf k}^{4})$,
\begin{eqnarray}
\bar{s}({\bf k}) &=& {1\over{2}}\, c\, (\ell^{2}{\bf k}^{2})^{2} 
+ O(|{\bf k}|^{6}).
\end{eqnarray}
For the Laughlin wave functions
\begin{eqnarray}
 c &=& (1-\nu)/(4\nu),
\label{cvalue}
\end{eqnarray}
where $\nu = 2\pi \ell^{2} \rho_{0}$.
The oscillator strength~\cite{GMP} 
\begin{eqnarray}
\bar{f}({\bf k}) &=& 2 \sum_{\bf p} V_{\bf p}\,
\Big(\sin {{\bf p\! \times\! k}\over{2}}\, \Big)^{2}\ \nonumber \\
&&\times  
 \Big[\, e^{{\bf - p\cdot k}}\, \bar{s} ({\bf p\!-\!k}) - 
e^{-{1\over{2}}\,{\bf k}^{2}}\, \bar{s} ({\bf p}) \Big]
\end{eqnarray}
also starts with $O({\bf k}^{4})$,
\begin{eqnarray}
\bar{f}({\bf k}) &=& 
{1\over{2}}\, \bar{\kappa}\, (\ell^{2}{\bf k}^{2})^{2} 
+ O(|{\bf k}|^{6}),
\end{eqnarray}
so that the excitation has a gap 
$\epsilon_{\rm SMA}^{\rm coll} = \bar{\kappa}/c$ at ${\bf k}=0$. 
The coefficient $\bar{\kappa}$ is given by
\begin{eqnarray}
\bar{\kappa}\! &=&\! \sum_{\bf p} V_{\bf p}\,\Big[  {{\bf
p}^{2}\over{2}}\, (D+{1\over{2}}) + {({\bf p}^{2})^{2}\over{4}}\,
(D+{1\over{2}})^{2} \,  \Big] \bar{s}[{\bf p}^{2}],\ \ \ \ \ \ \ \ 
\end{eqnarray}
where  $D=d/d({\bf p}^{2})$ acting on 
$\bar{s} ({\bf p})=\bar{s} [{\bf p}^{2}]$.

Let us now turn on $A_{\mu}$ and, as before, calculate 
terms that contribute to the $O(A^{2})$ response eventually.
See the Appendix for the evaluation of 
$i \langle\, [\theta\! \cdot\! \rho\ , 
\triangle H^{\rm Coul}] \,\rangle$
and 
$\langle\, \{\lambda\! \cdot\! \rho\ , 
\triangle H^{\rm Coul}\}\,\rangle$.
The result is 
\begin{eqnarray}
{\cal L}_{A}\!
&=&\!  \rho_{0} \Big[
-2\, \lambda\,\bar{s}\, \chi
- \bar{\kappa}\,\bar{\theta}\,{\bf k}^{2} \partial_{j} A_{j} 
+ \bar{\kappa}' \lambda \, {\bf k}^{2}A_{12} \Big]\ \ \ \ 
\label{LDLem}
\end{eqnarray}
to $O(\nabla^{3}A)$,
where 
$\bar{\kappa}'=\bar{\kappa} 
+ (1/2)  \sum_{\bf p} V_{\bf p}\,({\bf p}^{2})^{2}
D^{2} \bar{s}[{\bf p}^{2}]$.

The effective Lagrangian ${\cal L}_{0} +{\cal L}_{A}$, governing 
the collective mode $\bar{\theta}$, takes essentially the same
form as ${\cal L}_{A\theta}$ in Eq.~(\ref{LthetaA}).
[The effect of $\bar{\kappa}'\not =\bar{\kappa}$ is readily taken
care of by the replacement 
$\delta \rightarrow (\bar{\kappa}'/\bar{\kappa})\,
(1- s\omega_{c}/\bar{\kappa}')$ in the formulas of
Sec.~II, without spoiling gauge invariance.~\cite{fngaugeinv}] 
With appropriate rescaling of the fields it is seen that
the intra-Landau-level mode $\bar{\theta}$ behaves like 
a dipole-inactive $(0\rightarrow 2)$ cyclotron mode 
in the electron system of Eq.~(\ref{LthetaA}) with 
$\omega_{c}= \hat{f} \rightarrow 
{1\over{2}}\epsilon_{\rm SMA}^{\rm coll} \sim$ (Coulomb energy) 
at filling fraction $\nu _{\rm eff}=4\nu c=1-\nu$ 
(if we set $\hat{s}_{\rm eff} =1$).
Note that the collective mode disappears at $\nu =1$, as it should.

It is enlightening to compare the collective-mode spectrum with
that in the composite-fermion (CF) theory.
In the fermionic Chern-Simons theory of Lopez and Fradkin,~\cite{LF}
the random-phase approximation (RPA) around the mean field for
the Laughlin states with $\nu =1/3, 1/5,\cdots$ gives rise to 
a family of collective modes  with zero-momentum excitation gap
$q\,\omega_{\rm CF}$ and static structure factors 
$\bar{s} \sim ({\bf k}^{2})^{q}$, where
$q =2,3,\cdots, 1/\nu$, and  $\omega_{\rm CF} 
\equiv e B_{\rm eff}/M_{\rm CF} = \nu \omega_{c}$ stands for 
the Landau gap for composite fermions.  The mode with gap 
$\omega_{\rm CF}$ is missing and has been pushed up~\cite{LF,HLR} 
to the Landau gap $\omega_{c}$.
The lowest-lying collective mode, most stable among the family, 
thus has a gap $2\,\omega_{\rm CF}$ and spectral weight
$\bar{s} \sim ({\bf k}^{2})^{2}$.
Note that it has the same quadrupole character $\bar{s}
\sim ({\bf k}^{2})^{2}$ as the SMA collective mode.  It is therefore
natural to identify them and set
\begin{eqnarray} 
\epsilon^{\rm coll}_{\rm SMA} \approx 2 \omega_{\rm CF}\ \ 
{\rm for\ } {\bf k} \sim 0.
\label{SMAvswCF}
\end{eqnarray}
Let us here recall that in the CF theory, as discussed by
Goldhaber and Jain,~\cite{GJain} the composite fermions themselves
represent Laughlin's quasiparticles or vortices with 
fractional (renormalized) charge $-\nu e$ (and bare charge $-e$) 
and that $\omega_{\rm CF}$ is equal to the activation energy to 
create a widely-separated vortex-antivortex pair.~\cite{HLR,KH}
The spectrum~(\ref{SMAvswCF}) then suggests that the SMA collective
mode at ${\bf k}\!\sim\! 0$ consists of four vortices (or a two-roton
bound state~\cite{GMP} with the roton regarded as a vortex-antivortex
pair) in a quadrupole configuration so that
$\bar{s} \sim ({\bf k}^{2})^{2}$; this is in support of the Lee-Zhang
picture~\cite{LZ} of the magnetoroton branch at ${\bf k}\! \sim\! 0$ 
within the composite-boson CS theory.
This in turn gives, using the SMA value~\cite{GMP} for
$\epsilon^{\rm coll}_{\rm SMA}$, the activation energy
$\omega_{\rm CF}= {\scriptstyle {1\over{2}}}\, 
\epsilon^{\rm coll}_{\rm SMA} 
\approx 0.075\, (e^{2}/4\pi \epsilon^{*} \ell)$ for the $\nu = 1/3$
state, which is in rough agreement with other earlier
estimates.~\cite{HR,GMP} 
Furthermore, for the $\nu =1/5$ state the SMA estimate yields 
${\scriptstyle {1\over{2}}}\, \epsilon^{\rm coll}_{\rm SMA} 
\approx 0.025\, (e^{2}/4\pi \epsilon^{*} \ell)$ so that 
$(\omega_{\rm CF})_{\nu =1/5}/ (\omega_{\rm CF})_{\nu =1/3}
\approx 1/3$, 
which is not far from the ratio $(3/5)^{2}\approx 1/2.8$ expected
from the naive $\nu$ dependence of the activation energy.

The identification~(\ref{SMAvswCF}) has revealed
nontrivial consistency among the SMA theory and CS theories, both bosonic
and fermionic.  We remark that this is a nonperturbative yet
general result, in spite of the fact that in the CF theory corrections
beyond the RPA affect~\cite{LF,HLR} the activation gap
$\omega_{\rm CF}$ and the strength of higher-multipole responses
substantially. Note that Eq.~(\ref{SMAvswCF}) essentially follows from
the {\em absence} of a collective mode with a zero-momentum gap $\sim
\omega_{\rm CF}$, which is generally the case (otherwise, the
lowlying collective mode would become dipole-active, in violation of the
f-sum rule). It would thus hold for the (exact) renormalized
gap $\omega_{\rm CF}$ $\sim O(e^{2}/4\pi \epsilon^{*} \ell)$.

\section{Bilayer systems}

In bilayer systems, unlike single-layer systems, 
some of intra-Landau-level excitations become dipole-active, 
and this makes the SMA and CS theories more distinct.~\cite{MZ, KSdl}
In this section we construct an effective theory for bilayer systems.
For clarity of discussion we consider systems without interlayer
coherence and tunneling.

Consider a bilayer system with  average electron densities 
$\rho_{0}^{[\alpha]}=(\rho_{0}^{[1]},\rho_{0}^{[2]})$ 
in the upper $(\alpha =1)$ and lower $(\alpha =2)$ layers.
The system is placed in a common strong perpendicular magnetic field
$B$ and, as before, we focus on the lowest Landau level $n=0$ 
(with the electron fields $\psi^{[\alpha]}$ in each layer taken to be
fully spin polarized). 
The projected one-body Hamiltonian then reads
\begin{eqnarray}
\bar{H}_{1} &=& \sum_{\bf p}\{ \chi^{+}_{\bf p} \,
\bar{\rho}_{\bf -p} + \chi^{-}_{\bf p}\, \bar{d}_{\bf -p}\} ,
\end{eqnarray}
where $\bar{\rho}_{\bf p}= \bar{\rho}^{[1]}_{\bf p} +
\bar{\rho}^{[2]}_{\bf p}$ and 
$\bar{d}_{\bf p}= \bar{\rho}^{[1]}_{\bf p} -\bar{\rho}^{[2]}_{\bf p}$ 
are the projected charges; 
$\chi_{\bf p}^{\pm} =(A_{0}^{\pm})_{\bf p} +
(1/2M)({A}_{12}^{\pm})_{\bf p}$ and
$A_{\mu}^{\pm}(x)= {1\over{2}}\,\{A^{[1]}_{\mu}(x) \pm
A^{[2]}_{\mu}(x)\}$
in terms of weak external potentials $A^{[\alpha]}_{\mu}(x)$
acting on each layer.

The electrons in the two layers are coupled through the intralayer 
and interlayer Coulomb potentials $V^{11}_{\bf p}=V^{22}_{\bf p}$ 
and $V^{12}_{\bf p} = V^{21}_{\bf p}$, respectively; 
$V^{11}_{\bf p}= e^{2}/(2\epsilon^{*}\, |{\bf p}|)$ and 
$V^{12}_{\bf p} =e^{-d |{\bf p}|}V^{11}_{\bf p}$ with
the layer separation $d$ and the dielectric constant $\epsilon^{*}$ 
of the substrate. 
The projected Coulomb interaction is written as~\cite{fnbg} 
\begin{eqnarray}
\bar{H}^{\rm C}
={1\over{2}} \sum_{\bf p} \{ V^{+}_{\bf p}
\bar{\rho}_{\bf -p}\,\bar{\rho}_{\bf p} +
V^{-}_{\bf p} \bar{d}_{\bf -p}\, \bar{d}_{\bf p} \}
+ \triangle \bar{H}^{\rm C}
\label{projCoul}
\end{eqnarray}
with  $V^{\pm}_{\bf p}
={1\over{2}}(V^{11}_{\bf p} \pm V^{12}_{\bf p})$; 
the field-dependent piece $\triangle \bar{H}^{\rm C}$ takes
essentially the same form as in the single-layer case.

There is a variety of quantum Hall states in bilayer
systems.~\cite{Chak,BIH}  For definiteness we consider bilayer quantum
Hall states in a balanced configuration 
($\rho_{0}^{[1]}= \rho_{0}^{[2]}$), invariant under an interchange of 
the two layers.  Of our particular concern are bilayer states with
electron correlations, as described by  Halperin's $(m,m,n)$ wave
functions~\cite{BIH} at filling fractions $\nu= 2/(m+n)$; of these the
$(3,3,1)$ state at $\nu= 1/2$  has been observed
experimentally.~\cite{SES}

We here consider two types of collective excitations over such a
bilayer state $|G\rangle$, the in-phase density excitations of 
the two layers, $\bar{\rho}_{\bf k}|G\rangle$ (probed by $A_{0}^{+}$),
and the out-of-phase density excitations, $\bar{d}_{\bf k}|G\rangle$ 
(probed by $A_{0}^{-}$).
Note that $\langle \bar{\rho}_{\bf -k}\, \bar{d}_{\bf k}\rangle =0$ 
for balanced configurations.

Kohn's theorem~\cite{Kohn} implies that the in-phase collective
excitations remain dipole-inactive, as in the single-layer case, 
so that 
$\bar{s}_{+}({\bf k}) \sim |{\bf k}|^{4}$ for small ${\bf k}$. 
On the other hand, interlayer interactions $V^{12}_{\bf p}$ spoils
invariance under relative translations of the two layers and, 
unless interlayer coherence is realized, the out-of-phase collective
excitations become dipole-active,~\cite{MZ,RR}
\begin{equation}
\bar{s}_{-}({\bf k}) \equiv {1\over{N_{e}}}\, 
\langle \bar{d}_{\bf -k} \bar{d}_{\bf k} \rangle =
\hat{s}_{-}\,{1\over{2}}\,{\bf k}^{2} +  O(|{\bf k}|^{4}).
\label{sminus}
\end{equation}
For the $(m,m,n)$ states the coefficient $\hat{s}_{-}$ is given
by~\cite{RR}
\begin{eqnarray}
\hat{s}_{-} = 2n/(m-n).
\label{cminus}
\end{eqnarray}

To construct an effective theory let us denote the variations in phase
and amplitude of the in-phase mode by $\theta$ and $\lambda$, 
and those of the out-of-phase mode by $\xi$ and $\eta$.
Replacing $\theta \cdot\! \rho$ by  $\theta \cdot \!\bar{\rho} 
+ \xi \cdot \bar{d}$ and $\lambda \cdot \rho$ 
by  $\lambda \cdot\! \bar{\rho} + \eta \cdot \bar{d}$ 
in the single-layer expression~(\ref{Lthetalambda}) then yields an
effective Lagrangian. 
The result splits into the $(\theta,\lambda, A_{\mu}^{+})$ and
$(\xi,\eta, A_{\mu}^{-})$ sectors if  one, as before, only retains 
terms contributing to the $O(A^{2})$ response.

For the in-phase mode the effective theory is essentially the
same as the single-layer case  ${\cal L}_{0}+{\cal L}_{A}$ in
Eqs.~(\ref{LDL}) and~(\ref{LDLem}) with 
$A_{\mu} \rightarrow A_{\mu}^{+}$, $\bar{\kappa} 
\rightarrow \bar{\kappa}^{+} +\bar{\kappa}^{-}$ and 
$\bar{\kappa}' \rightarrow (\bar{\kappa}')^{+} + (\bar{\kappa}')^{-}$.
Here $\bar{\kappa}^{\pm}$ and $(\bar{\kappa}')^{\pm}$ are given by the
corresponding single-layer expressions with 
$[V^{\pm}_{\bf p},\bar{s}^{\pm}({\bf p}) ]$.

For the out-of-phase mode the oscillator strength~\cite{MZ,RR} starts 
with ${\bf k}^{2}$, 
\begin{eqnarray}
&&f_{-}({\bf k}) 
= {1\over{2}}\, \kappa_{-}\, {\bf k}^{2} +  O(|{\bf k}|^{4}) ,
\nonumber\\ 
&&\kappa_{-} = 2\sum_{\bf p} {\bf p}^{2}\, V^{12}_{\bf p}
\{-\bar{s}^{12}({\bf p}) \} ,
\end{eqnarray}
where $ \bar{s}^{12} ({\bf p}) \equiv 
\langle G|\bar{\rho}^{[1]}_{\bf -p}\,\bar{\rho}^{[2]}_{\bf p}\, 
|G\rangle /N_{e} 
={1\over{4}}\, \{ \bar{s}_{+} ({\bf p}) -\bar{s}_{-} ({\bf p})\}$.  
This leads to the SMA excitation gap 
$\epsilon^{\rm coll}_{-} = \kappa_{-}/\hat{s}_{-}$ at 
${\bf k}\rightarrow 0$.
Eventually one is led to an effective Lagrangian of the form 
\begin{eqnarray}
{\cal L}^{\rm coll}_{-} = \rho_{0}\, 
\Big[&& \!\!\!\!\!\! 2\eta\,s_{-}\, (\dot{\xi}- \chi^{-})
 - (\eta\, f_{-} \,\eta + \xi\,f_{-} \,\xi) 
\nonumber\\
&& - \kappa_{-}\, \xi\, \partial_{j}A_{j}^{-} 
+ \kappa_{-} \, \eta\,A^{-}_{12} \Big],
\label{LeffDL}
\end{eqnarray}
apart from terms of $O(\partial^{3}A^{-})$.
Again the coefficient $\kappa_{-}$ of 
the $\xi\, \partial_{j}A_{j}^{-}$ term is correlated with 
$f_{-}({\bf k})$, 
in conformity with gauge invariance.~\cite{fngaugeinv} 
With obvious substitution this ${\cal L}^{\rm coll}_{-}$ is cast into 
the form of the effective Chern-Simons theory and dual-field theory 
of Eqs.~(\ref{Leffcs}) and (\ref{Lb}), respectively, and leads to 
an out-of-phase response of the form of Eq.~(\ref{emresponse}).

On the other hand, the $(0\rightarrow 1)$ cyclotron modes
$\xi^{[\alpha]}$ associated with each layer $\alpha =1,2$ are
described by the effective CS theory of Eq.~(\ref{Leffcs}) 
with $\hat{s} \rightarrow 1$ and $\xi \rightarrow \xi^{[\alpha]}$, 
$A_{\mu} \rightarrow A_{\mu}^{[\alpha]}$,
$\rho_{0} \rightarrow \rho_{0}^{[\alpha]}$, etc.
The intra and interlayer Coulomb interactions are also correctly
incorporated by use of an appropriate Hubbard-Stratonovich
transformation.  The effective theory agrees with the standard
bilayer bosonic CS theory,~\cite{WZdlayer,EI} except that the CS term 
has no interlayer mixing component. 
For the $(m,m,n)$ states the relevant mixing matrices 
(in terms of $\hat{s}$) differ by  
\begin{eqnarray}
 \left(\begin{array}{cc}
1  & 0 \\
 0   & 1
           \end{array} \right)
 \leftrightarrow 
{1\over{m - n}} \left(\begin{array}{cc}
m & -n\\
-n   & m
           \end{array} \right) .
\label{mixingmatrix}
\end{eqnarray}
The latter matrix in the CS theory is diagonalized in the
$(\bar{\rho},\bar{d})$ basis, yielding
$\hat{s}^{\rm CS}_{+} =1$ (hence the correct Hall conductance 
$\nu e^{2}/h$) for the in-phase cyclotron mode and 
$\hat{s}^{\rm CS}_{-} =(m+n)/(m-n)$ for the out-of-phase mode.

It is important to note here that these dipole components of 
the structure factors govern the long-wavelength structure of
many-body wave functions, as pointed out 
by Lopez and Fradkin.~\cite{LFdl}
This implies, in particular, that, if the quantum Hall state $|G\rangle$
embodies electronic correlations characteristic of the $(m,m,n)$ wave
functions, one must have 
\begin{eqnarray}
\hat{s}^{(mmn)}_{+} =1 \ {\rm and}\ \ 
\hat{s}^{(mmn)}_{-} =(m+n)/(m-n).
\label{mmncorrelation}
\end{eqnarray}
This condition cannot be fulfilled by the cyclotron modes alone, which
yield $\hat{s}^{(10)}_{\pm}= 1$, an inevitable consequence of the 
(projected) $f$-sum rule. 
This, in turn, implies the necessity of dipole-active
intra-Landau-level excitations.  
Indeed, as seen form the SMA response~(\ref{emresponse}), 
the cyclotron mode and collective mode combine to yield the desired 
out-of-phase $(m,m,n)$ electronic correlations
\begin{eqnarray}
1+\hat{s}_{-} = {m+n\over{m-n}} =\hat{s}^{(mmn)}_{-}.
\label{smmn}
\end{eqnarray}

On the other hand, the bilayer bosonic CS theory saturates 
the (out-of-phase) $f$-sum rule by a single dipole-active mode
so that 
$\hat{s}^{(mmn)}_{-}\omega^{\rm coll}_{\rm CS}\approx \omega_{c}$.
One thereby finds its spectrum at
\begin{equation}
\omega^{\rm coll}_{\rm CS}=[(m-n)/(m+n)]\,\omega_{c}
\end{equation} 
for ${\bf k}\rightarrow 0$.  This unnatural shift of the out-of-phase
cyclotron mode is attributed to the lack of projection in the bosonic 
CS theory (which thus fails to distinguish between the intra- and
inter-Landau-level modes).

In the bilayer fermionic CS theory~\cite{LFdl} there emerge two
dipole-active modes with strength ${1\over{2}}\,\hat{s}^{(mmn)}_{-}$ 
at $\omega^{\rm coll}_{\rm CS}$, thus resulting in essentially 
the same situation as in the bosonic CS theory.
Apart from the collective-excitation spectrum, however, the SMA theory
reproduces, owing to Eq.~(\ref{smmn}), the favorable long-wavelength
transport properties of the CS theories, such as the Hall conductance,
long-range orders, and fractional vortex charges.

A good place to detect the out-of-phase collective mode would be 
Hall drag experiments.~\cite{KSEPW}
The interlayer Hall conductance becomes sizable~\cite{Renn} in the
presence of the $(m,m,n)$ correlations, as read from the
electromagnetic response~(\ref{emresponse}): 
\begin{eqnarray}
 \left(\begin{array}{c}
\!\! J_{x}^{[1]} \!\!\\
\!\! J_{x}^{[2]} \!\!
           \end{array} \right)
= -{e^{2}\nu \over{4h}}
 \left( \! \begin{array}{cc}
\sigma^{+}\! + \sigma^{-} &\sigma^{+}\! - \sigma^{-}\\
\sigma^{+}\! - \sigma^{-} &\sigma^{+}\! + \sigma^{-}
           \end{array} \! \right) 
 \left(\begin{array}{c}
\!\! E_{y}^{[1]} \!\!\\
\!\! E_{y}^{[2]} \!\!
           \end{array} \right),
\label{mixingmatrix}
\end{eqnarray}
where $\sigma^{+}= 1/[1 - (\omega/\omega_{c})^{2}]$ and 
$\sigma^{-}= 1/[1 - (\omega/\omega_{c})^{2}]
+\hat{s}_{-}/[1 - (\omega/\epsilon_{-}^{\rm coll})^{2}]$.
For the (3,3,1) state a direct current $J_{x}^{[1]}$ injected to the upper
layer would induce a Hall voltage 
$V_{y}^{[2]} = (h/e^{2}) J_{x}^{[1]}$ on the lower layer 
(left to be an open circuit) while yielding
$V_{y}^{[1]} = 3\, (h/e^{2}) J_{x}^{[1]}$ on the same layer. 
The interlayer resistivity is sensitive to the collective mode 
through its response to time-varying currents, 
\begin{equation}
\rho_{yx}^{[21]} \approx - (h/e^{2})\, n 
/ [ 1-\beta\, \omega^{2}/(\epsilon_{-}^{\rm coll})^{2}]
\end{equation} 
with $\beta=(m-n)/(m+n)$ while 
$\rho_{yx}^{[11]}+ \rho_{yx}^{[21]} \approx  -(h/e^{2})\,
(m+n)$ stays fixed for 
$\omega \le \epsilon_{-}^{\rm coll} \ll \omega_{c}$. 
Since the quantized Hall resistance (for $\omega =0)$ is expected to be
very accurate, there would be a good chance of detecting such a
difference in experiments with an injected high-frequency current or a
current pulse.

\section{Summary and discussion}

In this paper we have presented a unified treatment of elementary and
collective excitations in quantum Hall systems by means of the
single-mode approximation (SMA).  A variational principle and
nonlinear realizations of the $W_{\infty}$ algebra have been combined
to construct effective theories that incorporate the SMA excitation
spectrum. 
For a wide class of quantum Hall systems the resulting effective
theory turns out to be a variant of the bosonic CS theory.  
In this sense, there is a direct link between the composite-boson 
theory and the SMA treatment of quantum Hall systems. 
The CS action therefore is a quite natural element for quantum Hall
systems, irrespective of the notion  of flux attachment.

We have noted that the intra-Landau-level modes bear a similarity 
in structure (though not in scale) to the inter-Landau-level modes.
This has revealed a further link with CS theories:
A comparison with the fermionic CS theory suggests that
the SMA collective modes in single-layer systems 
(around ${\bf k} \approx 0$) are composed of four vortices 
in a quadrupole configuration, in conformity with 
an interpretation~\cite{LZ} within the bosonic CS theory.

In the SMA all the information is essentially contained in 
the structure factors $s({\bf k})$, which represent electronic
correlations pertinent to the quantum Hall state in question.   
In particular, the dipole part $\propto {\bf k}^{2}$ of $s({\bf k})$ 
is directly related to the long-wavelength structure of 
many-body wave functions and also to the flux attachment
transformation employed in CS approaches.  
This fact has some important consequences. First, for single-layer
systems such $O({\bf k}^{2})$ correlations are governed by the
cyclotron modes alone (as implied by Kohn's theorem), so is the
CS-flux attachment.
This resolves the puzzle why the main features of the single-layer 
CS theories, such as the Hall conductance, long-range orders and
fractional charges of quasiparticles, are apparently determined by 
the cyclotron modes (or the electronic kinetic term which one would
naively expect to be quenched)  while the FQHE is actually caused 
by the Coulomb interaction.   
In this sense, the dipole correlation with $\hat{s}^{(10)}= 1$ is
characteristic of typical single-layer quantum Hall states.

For bilayer systems the situation is different.
Correlations characteristic of the bilayer $(m,m,n)$ states, 
for example, are not attained by the cyclotron modes alone and require
the presence of dipole-active collective modes.  
As a result, the SMA effective theory is bound to involve such
collective modes and deviates from a naive bilayer version of 
the bosonic CS theory.

While the structure factors $s({\bf k})$ are readily calculated 
for the cyclotron modes via projection, the determination of 
$s({\bf k})$ for collective modes is a nontrivial subject of dynamics, 
handled in a variety of approximation schemes. 
In the present paper we have simply focused on quantum Hall states
well approximated by Laughlin's or Halperin's wave functions. 
There is a practical way to improve the structure factors. 
One may appeal to Jain's composite-fermion wave-function
approach,~\cite{J} which is known to yield numerically very accurate
variational wave functions and which, along with refinement 
to incorporate corrections due to layer thickness, Coulombic
Landau-level mixing, etc., has been generalized to bilayer systems 
as well.~\cite{SPJ}   
With such improved structure factors one could refine the effective
theory and make comparisons with experiments more reliable.

\acknowledgments

This work was supported in part by a Grant-in-Aid for Scientific
Research from the Ministry of Education of Japan, Science and Culture
(Grant No. 14540261).

\appendix

\section{calculation}
In this appendix we outline the calculation of the electromagnetic
coupling~(\ref{LDLem}) coming from the field-dependent
Coulomb interaction $\triangle H^{\rm Coul}$.   
Consider a factor 
$\Lambda_{\bf k,p}= (1/2N_{e})\, 
\langle G| \bar{\rho}_{\bf k}\,  \{ \bar{\rho}_{\bf -p}\, ,
\,\bar{\rho}_{\bf p\! -\!k}\, \} |G \rangle$
involving products of three projected charges.  
As discussed in an earlier SMA treatment,~\cite{KSdl} the leading
small ${\bf k}$ behavior of such products is determined from the
portion that originates from the noncommutative nature 
$[r_{1}, r_{2}] = i\ell^{2}$ of ${\bf r}$, 
with the result
\begin{eqnarray}
 \Lambda_{\bf k,p}=\!\! &&  (e^{-{1\over{2}} k^{\dag}p} -1 )\,
\bar{s}({\bf p\!-\!k})  +   (e^{{1\over{2}} k^{\dag}(p-k)} -1 )\, 
\bar{s}({\bf p})
\nonumber\\
&&  + O({\bf k}^{4}),
\end{eqnarray}
where $k^{\dag}p= {\bf k\cdot p} -i{\bf k \times p}$.
The factors relevant for the
$\langle\,\{\lambda\! \cdot\! \rho\ , \triangle H^{\rm Coul}\}\,\rangle$
and
$i \langle\, [\theta\! \cdot\! \rho\ , \triangle H^{\rm Coul}]\,\rangle$
terms are then given by the real and imaginary part 
of $\Lambda_{\bf k,p}$, respectively.  
Note, as an independent check, that 
$[\theta\cdot\rho , \triangle H^{\rm Coul}]$ is also determined
from the charge algebra~(\ref{chargealgebra}) alone, yielding 
the same result.


\begin{thebibliography}{99}
\bibitem{fngaugeinv}
Also a direct calculation shows that a gauge-variant $O(A^{2})$ term
following from Eq.~(\ref{LDLem}) is canceled by a term coming from 
the field-dependent Coulomb interaction of $O(A^{2})$; analogously 
for Eq.~(\ref{LeffDL}).
\end{thebibliography}
\end{document}